\documentclass[pre,twocolumn]{revtex4}
\usepackage{amsmath}
\usepackage{amsfonts}
\usepackage{graphicx}
\usepackage{amssymb}

\begin{document}

\title{
Enhanced Septahedral Ordering in Cold Lennard-Jones Fluids
 }

\author{Paul C. Whitford}

\affiliation{Department of Physics, University of California, San Diego, La Jolla, CA 92093}

\author{George D. J. Phillies}
\email{phillies@wpi.edu}

\affiliation{Department of Physics, Worcester Polytechnic
Institute,Worcester, MA 01609}

\begin{abstract}

We report molecular dynamics calculations on two-component, cold ($1.2 \geq T \geq 0.56$ in natural units), three-dimensional Lennard-Jones fluids.  Our system was small (7813 A, 7812 B particles), dense ($N/V = 1.30$), and distinctly finite ($L \times L \times L$ cube, periodic boundary conditions, with $L=22.96 \sigma_{\rm AA}$), $\sigma_{\rm AA}$ being the range of the $AA$ interaction in the Lennard-Jones potential
$U_{ij} = 4 \epsilon \left[ \left( \frac{\sigma_{ij}}{r}\right)^{12} -\left( \frac{\sigma_{ij}}{r}\right)^{6}\right]$.  We calculated spherical harmonic components $Q_{\rm LM}$ for the density of particles in the first coordination shell of each particle, as well as their spherical invariants $\langle Q_{\rm L}^{2} \rangle$, time-correlation functions and wavelet density decompositions.  The spherical invariants show that non-crystalline septahedral ($\langle Q_{7}^{2} \rangle$) ordering is important, especially at low temperature. While $\langle Q_{10}^{2} \rangle$ could arise from icosahedral ordering, its behavior so closely tracks that of the nonicosahedral  $\langle Q_{11}^{2} \rangle$ that alternative origins for $\langle Q_{10}^{2} \rangle$ need to be considered.  Time correlation functions of spherical harmonic components are bimodal, with a faster temperature-independent mode and a slow, strongly temperature-dependent mode.  Microviscosities inferred from mean-square particle displacements are exponential in static amplitude $\langle Q_{7}^{2} \rangle$, and track closely in temperature dependence the orientation density slow mode lifetime.  Volume wavelet decompositions show that when $T$ is reduced, the correlation length of $\langle Q_{7}^{2} \rangle$ increases, especially below $T=0.7$, but the correlation length of $\langle Q_{5}^{2} \rangle$ is independent of $T$.
 
\end{abstract}

\maketitle

\section{Introduction}

Many fluid systems, on cooling below their freezing points, decline to crystallize, at least on experimentally accessible time scales.  The fluids instead become glasses, highly viscous liquids or amorphous solids that appear rigid, but that lack long range crystalline order.  Despite enormous experimental, theoretical, and simulational study, the nature of the glass transition remains unclear.  Is the transition toward the glass a dynamic effect, or does it reflect thermodynamic issues?  A vast range of possible physical quantities  might be correlated with glass formation, only a few of which have been studied systematically. 

This paper treats orientational order, and fluctuations in orientational ordering, in cold Lennard-Jones fluids.  Our results are based on molecular dynamics simulations.  Orientational ordering in glass-forming liquids and other systems has been studied extensively\cite{steinhardt1983a,jonsson1988a,terrones1994a,sanyal1995a,tomida1995a,tanaka2003a,cicco2003a,jakse2003a,tanaka2000a,wang1991a}.  The work reported here differs from earlier studies in that we (i) compute orientation correlations that correspond to non-crystalline ordering, (2) perform wavelet decompositions of the orientation correlation density, and (3) determine the temporal evolution of fluctuations in orientational ordering.

There are a variety of ways to characterize angular order around a given particle.  For example, one can measure the 'bond angle distribution', the relative likelihood that the displacement vectors from the given atom to two neighboring atoms have a particular value\cite{jakse2003a}.  The bond angle distribution, which reflects an aspect of the three-particle distribution function $g^{(3)}({\bf r}_{1}, {\bf r}_{2}, {\bf r}_{3})$, has recently been shown to be accessible to experimental measurement\cite{cicco2003a}.  Steinhardt, et al.\cite{steinhardt1983a} used a spherical harmonic decomposition of the near neighbors of a given particle, an idea referred back to the earlier work of Frank\cite{frank1952a}.  Spherical harmonics are a complete set of functions over the sphere, so the angular distribution of (for example) an atom's nearest neighbors can always be expanded in spherical harmonics of their density.

To characterize an angular distribution, one chooses an atom of interest, and a set of other atoms whose angular distribution around the atom of interest is to be characterized.  The unit vector $\hat{\bf r}_{i}$ from the atom of interest to each atom $i$ of the $N$ other atoms are generated.  The LM$^{\rm th}$ spherical harmonic component $Q_{\rm LM}$ of the $\hat{\bf r}_{i}$ is
\begin{equation}
     Q_{\rm LM} = N^{-1} \sum_{i=1}^{N} Y_{\rm LM}(\hat{\bf r}_{i})
     \label{eq:QLMdef}
\end{equation}
where $Y_{\rm LM}$ is the LM$^{\rm th}$ spherical harmonic, and in the spherical harmonic  $\hat{\bf r}_{i}$ is represented by the polar angles $(\theta, \phi)$.  The polar angles refer to a choice of the angular origins.  To eliminate the dependence of $Q_{\rm LM}$ on this nonphysical feature, one forms spherical invariants
\begin{equation}
   Q_{\ell}^{2} = \frac{4 \pi}{2 \ell + 1}\sum_{m = -\ell}^{\ell}  \mid Q_{\ell m} \mid^{2}
   \label{eqQLdef}
\end{equation}
that are independent of coordinate axes\cite{steinhardt1983a}.    

Many prior treatments of particle orientations focused on a search for icosahedral, cubic, or other crystalline local ordering in the fluid, including local orderings that are not crystallographic because they do not lead to space-filling packings.  Partial crystalline orderings are characterized by particular values for some of the $Q_{\ell}^{2}$.  For example, icosahedral order around a center leads to nonzero $Q_{l}^{2}$ for $\ell = 6, 10, 12$ while face-centered-cubic and body-centered-cubic clusters have $Q_{\ell}^{2} \neq 0$ for $\ell = 4, 6, 8$.
As applied to liquids, the ensemble-average $\langle Q_{\ell}^{2} \rangle $ determined from simulation or elsewise are compared with the $\langle Q_{\ell}^{2} \rangle $ that would arise from ideal local order of one sort or another.
For historical reasons, prior work has largely been limited to $\langle Q_{\ell}^{2} \rangle$ for even $\ell$.  However, the $Q_{\ell}^{2}$ are positive semidefinite, and fluctuations can certainly lead to non-zero values for any $Q_{\ell}$, so therefore it is inescapably the case that $\langle Q_{\ell}^{2} \rangle$ is  nonzero for every $\ell$.

The spherical invariant $Q_{\ell}^{2}$ is determined by the instantaneous particle positions, and is therefore a function of time and space: $Q_{\ell}^{2}$ changes as particles that are initially nearby move or are replaced with other particles.  At a given moment, $Q_{\ell}^{2}$ also varies from atom to atom.  Wang and Stroud\cite{wang1991a} report studies of a spatial correlation function $\langle Q_{\rm LM}({\bf r_{a}})Q^{*}_{\rm LM}({\bf r_{a}}+ {\bf r}) \rangle $for the spherical harmonic components around two atoms separated by ${\bf r}$.  The time correlation function of the time-dependent $Q_{\rm LM}(t)$, namely
\begin{equation}
   C_{\ell}^{(2)}(t) =\frac{4 \pi}{2 \ell + 1}\sum_{M = -\ell}^{\ell}  \langle Q_{\ell{\rm M}}(0)Q^{*}_{\ell{\rm M}}(t)\rangle
   \label{eq:c2tdef}
\end{equation}
describes the relaxation of orientational order fluctuations around a given particle.  Sanyal and Sood\cite{sanyal1995a} studied a variation on $C_{\ell}^{(2)}(t)$, in which an elaborate weighted sum of $Q_{\rm LM}$ over all particles in small cubic subsections of the system was taken.  The time correlation function of the weighted sum was computed for each cubelet. Sanyal and Sood's 'bond-orientation correlation function' had a relaxation that was fit adequately well by a stretched exponential in time.

We further treat wavelet decompositions of the  density of the $\langle Q_{\ell}^{2} \rangle$.
Wavelets\cite{strang} provide via functional transformation a representation of a function $g(x)$.  A wavelet transform $\tilde{g}(a,b)$ of a function $g(x)$ is obtained as
\begin{equation}
       \tilde{g}(a,b) = \int dx \ g(x) F(a,b;x)
       \label{eq:transform}
\end{equation}
where $a$ and $b$ are wavelet parameters, the \emph{dilation} and the \emph{translation}, respectively, and where the $F(a,b;x)$ are the basis vectors of the wavelet transformation.  The $\tilde{g}(a,b)$ are the wavelet components of the function $g(x)$.  The basis vectors all follow from a single mother wavelet function $f(x)$
via 
\begin{equation}
      F(a,b;x) = f(a x - b);
    \label{eq:families}
\end{equation}
the effectiveness of the transform being determined by the choices of $f$, $a$, and $b$.  One choice of dilation and transformation variables relies on binary decimation, namely $a = 2^{m}$ and $b = n \ell_{0} 2^{m}$, $\ell_{o}$ being a basic length and $n$ and $m$ being integers.  The $\tilde{g}(a,b)$ allow one to regenerate the original $g(x)$, namely
\begin{equation}
     g(x) = \int \ da \ db \ \tilde{g}(a, b) H(a, b; x).
     \label{eq:transformback}
\end{equation}
Here the integral (which may actually be a sum) covers all allowed $a$ and $b$, and the $H(a,b;x)$ are the vector duals of the $F(a,b;x)$.  

Wavelet transforms are superficially a generalization of Fourier transforms, for which the basis vectors are the $\exp(-i k x)$ and the dual vectors are the same function $\exp( i k x)$.  In most cases, for wavelet transforms the $F$ and the $H$ differ markedly.  Musical notation, now twelve centuries old, gives an early description of a continuous function using basic vectors--namely musical notes--that have local support, and that arise from translation and dilation. Equations \ref{eq:transform}-\ref{eq:transformback} assign to the function being transformed a continuous index $x$, but there is no mathematical restriction that $x$ be continuous.  It is equally possible to apply wavelet transforms to a function $g_{i}$ that is labelled by a discrete index $i$.

Wavelet and their transforms in general have a variety of important mathematical properties:
   
   (i) local support.  The support of a function $F$ is the region over which it is nonzero.  Most wavelets have only local support, and are non-zero over a limited range of $x$.  In contrast, the Fourier transform function $\exp(-ikx)$ has non-local support, and is non-zero almost everywhere.

    (ii) local reconstruction.  Changing a $\tilde{g}(a,b)$ changes the reconstructed function (eq \ref{eq:transformback}) over a limited region, namely the support of the associated $H(a,b;x)$.

     (iii) On the other hand, most wavelet basis vectors are not antisymmetric or symmetric under $x \rightarrow -x$.  Nor are they in general eigenfunctions of obvious differential operators.  Indeed, many wavelets only have a limited number of derivatives.  In some wavelet families, the $F(a, b;x)$ are not orthogonal; they are instead overcomplete (though reconstruction is still possible).

An extremely wide range of continuous and discrete wavelet transforms is known to exist.  Much of the published literature on wavelet transforms examines one-dimensional transforms.  To apply transforms to a volume one must either construct novel three-dimensional transforms, or one must construct a three-dimensional transform as the outer product of a series of one-dimensional transforms.  We followed the latter approach.  Space was divided into small volumes, a spherical harmonic density was computed for each volume, and discrete wavelet transforms were then performed. 

We applied the discrete Haar\cite{haar1910a} wavelets, which act by applying to adjoining pairs of points a low-pass filter $C$ and a highpass filter $D$. The filters act on an adjoining pair of points $x_{2n-1}$ and $x_{2n}$ as
\begin{equation}
     C_{n} = (x_{2n-1} +x_{2n})/2
     \label{eq:Cdef}
\end{equation}
and 
\begin{equation}
     D_{n} = (x_{2n-1} -x_{2n})/2.
     \label{eq:Cdef}
\end{equation}
The $C$ filter produces a smoothed value, and the $D$ filter acts as an edge detector.  The transform is done iteratively, by applying further pairs of Haar\cite{haar1910a} filters to the output of the $C$ filter from the previous iteration. Successive iterations are sensitive to features extending over larger spatial domains. To perform a Haar transform on a cubic array of points, one first applies the $C$ and $D$ filters to pairs of points along one axis.  The $C$ and $D$ filters are then applied to pairs of $1 \times 2$ blocks along the second axis, and then applied along the third axis to filtered $2 \times 2$ plates along a third axis, so that, e.g.,
\begin{equation}
     CCC_{111} = x_{111}+x_{211}+x_{121}+x_{112}+x_{221}+x_{122}+x_{212}+x_{222}
     \label{eq:CCCdef}
\end{equation}
while
\begin{equation}
     DDD_{111} = x_{111}-x_{211}-x_{121}-x_{112}+x_{221}+x_{122}+x_{212}-x_{222}.
     \label{eq:CCCdef}
\end{equation}
Here $x_{ijk}$ is the value of $x$ at the triple labelled point $ijk$.  Because fluids have no directional orientation, the various $CCD$, $CDC$, and $DCC$ filters should on thermal averaging give the same output.  The wavelet decomposition is iterated by applying further cycles of $C$ and $D$ filters to the $CCC$ output from the prior iteration, so that after each level of decomposition the $CCC$ outcomes are averaged over a cubical volume with twice the linear extent.  Physically, the $CCC$ filter detects the uniformity in $x$ across a region, the $CCD$ filters are sensitive to surfaces bifurcating a region into zones with different values for $x$, the $CDD$ filters are sensitive to lines where pairs of surfaces intersect, and the $DDD$ filter identifies vertices where trios of surfaces come together.  The filters other than $CCC$ give signed output, so we computed averages such as $\langle (CCD)^{2}\rangle$.

\section{Computational}

We studied a two-component A-B Lennard-Jones fluid.  The potential energy was
\begin{equation}
    U_{ij} = 4 \epsilon \left[ \left( \frac{\sigma_{ij}}{r}\right)^{12} -\left( \frac{\sigma_{ij}}{r}\right)^{6}
\right]
   \label{eq:uijform}
\end{equation}
Here $i$ and $j$ label the species ($A$, $B$) of the interacting particles, so that $N_{A}$ and $N_{B}$ are the number of $A$ and $B$ particles, while $r$ is the distance between centers of mass.  Parameters were taken from Glotzer, et al.\cite{glotzer2000a}, namely $\sigma_{AA} = 1$, $\sigma_{BB} = 5/6$, and $\sigma_{AB}=11/12$.   All particles had the same mass $m$ and energy constant $\epsilon$.  Natural units are used throughout, so the temperature $T$ is units of $k_{B}/\epsilon$, $m=1$ so force and acceleration are equal, and the time unit is $\sigma_{AA} (m/\epsilon)^{1/2}$.  The potential was truncated: $U_{ij}(r)=0$ for $r > 2.5$. The underlying molecular dynamics code is identical to that of our earlier paper\cite{whitford2005a}:  Newton's equations of motion were integrated numerically using the Calvo and Sanz-Serna fourth order method\cite{calvo1993a,gray1994a}, applying standard approaches to minimize calculation time.

Our system is smaller ($N_{A} = 7813$ as opposed to $N_{A} = 62,500$) than the system of our previous paper\cite{whitford2005a}.  We first did the calculations reported here and found that $g(r)$ has a much larger range than sometimes reported.  In order to determine $g(r)$ accurately, the 15,625 particle system of this paper was replaced with the 125,000 particle system on which we have published\cite{whitford2005a}. The A:B mixture was almost exactly 1:1; $N_{A}$ and $N_{B}$ differ by one atom. The density is the same as that of the previously-reported simulation: $N/V =1.30$, where $N$ is the total number of particles and $V= L^{3}$ is the system volume, with $L=22.96$.  Based on the range of known static correlations (at $T=0.56$, $g(r)-1 \neq 0$ out to $r \geq 10$), the fluid studied here is weakly periodically confined, in the sense that the radial distribution shells around two particles can overlap twice, once in each of two opposite directions.  

We equilibrated our systems at temperatures 1.20, 0.88, 0.73, 0.69, 0.66, 0.64, 0.59, and 0.56.  The intermediate temperatures were chosen to match Ref.\ \onlinecite{glotzer2000a}. To equilibrate the system, we prepared an initial system at $T=1.20$, equilibrated that system, and then cooled to new temperatures by rescaling the particle momentum by 0.1\% every fifty time steps, a time step being 0.01 in natural units.  On reaching the target temperature, the system was equilibrated for 10,000 time steps while monitoring the average kinetic energy. If the energy drifted during the 10,000 equilibration steps, the system was brought again to the target temperature and re-equilibrated until the average kinetic energy finally remained near its desired target value.  Simulations at a given temperature were then run for at least 50,000 time steps.

The generic radial distribution function $g(r)$ (which gives the normalized ($g(\infty) = 1$) probability of finding any two atoms a distance $r$ apart) and the specific radial distribution functions $g_{ij}(r)$ (which give the probability of finding two particles of species $i$ and $j$ a distance $r$ apart) were determined.  The mean-square particle displacements, the $\langle Q_{\ell}^{2}\rangle$, and the $C_{\ell}^{(2)}(t)$ were then computed, taking the particles neighboring the particle of interest to be the particles in the first peak (region where $g(r) > 1$) of $g(r)$, including $\ell \in (1,12)$.  Static correlation functions $\langle Q_{\ell}^{2}\rangle$ were evaluated using every particle as a center and repeating the calculation every fifty time steps.

We examined a wavelet decomposition of several spherical harmonic component densities. 
In order to apply wavelet decompositions, we needed a rule for assigning wavelet densities to points $ijk$.  We examined two rules, namely
\begin{equation}
  \rho_{ijk}^{(1)} = \frac{\sum_{V_{ijk}} \left( Q^{2}_{\ell} - \langle Q^{2}_{\ell}\rangle \right)}{V_{ijk}}
      \label{eq:rho1}
\end{equation}
and     
\begin{equation}
      \rho_{ijk}^{(2)} = \frac{ n_{ijk}^{-1} \sum_{V_{ijk}} \left( Q^{2}_{\ell} - \langle Q^{2}_{\ell}\rangle \right)}{V_{ijk} \sigma_{Q} }
      \label{eq:rho2}
\end{equation}
Here the system is divided into cubelets having volumes $V_{ijk}$, each containing $n_{ijk}$ atoms used as central particles for the evaluation of the $Q_{\ell}$ centered on them, and $\sigma_{Q}$ is the root mean square deviation of the distribution of the $Q_{l}$.  Equation \ref{eq:rho1} gives the total deviation within $V_{ijk}$ of $Q_{\ell}$, while eq \ref{eq:rho2} gives the deviation as averaged over all the atoms within $V_{ijk}$.  Further work here is based on eq \ref{eq:rho1}.

Finally, we examined a position-momentum time correlation function
\begin{equation}
    \langle Y_{1} P(\tau) \rangle = \frac{\langle Q_{10}(t) p_{z}(t+\tau) \rangle }{( \langle \mid Q_{10}(t)\mid^{2}\rangle  \langle p^{2}_{z}(t)\rangle   )^{1/2}}
    \label{eq:Y1Pdef}
\end{equation}
and a position-kinetic energy time correlation function
\begin{equation}
\langle Y_{2}K(\tau) \rangle= \frac{\langle Q_{20}(t) K_{z}(t+\tau) \rangle }{( \langle \mid Q_{20}(t)\mid^{2}\rangle  \langle K^{2}_{z}(t)\rangle   )^{1/2}}
    \label{eq:Y2Kdef}
\end{equation}
Here $p_{z}$ is the $z$-component of the momentum of the particle of interest, while $K_{z} = p_{z}^{2}/(2m)$, labels $(t)$ and $(t+\tau)$ refer to determinations on the same particle of interest at the indicated times, and the $\theta=0$ direction is taken to be the $z$-axis.  For delay time $t=0$, both $\langle Y_{1} P(t) \rangle$ and $\langle Y_{2}K(t) \rangle$ must vanish, because equal-time position-momentum correlations cannot exist in classical systems.  However, in each correlation function the $Y_{\rm LM}$, $p_{i}$, and $K_{i}$ were chosen so that they had exactly matching spatial tensor properties, so that the correlations in their products were not forced to zero by reflection or another spatial symmetry. For $t \neq 0$, we found that these correlations have a positive- or negative-going onset, respectively.

\section{Results}

\begin{figure}[thb]

\includegraphics{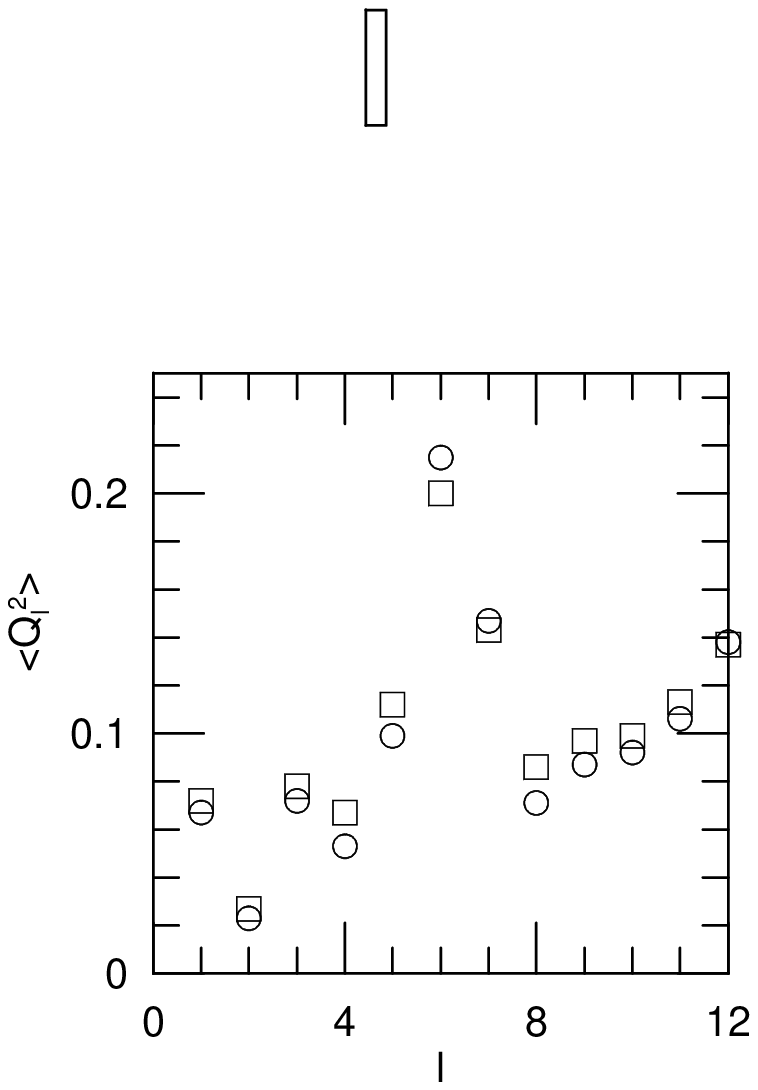}

\caption{\label{figurezero}  Spherical invariants $\langle Q_{\ell}^{2} \rangle$ as a function of $\ell$ at temperatures 0.55 ($\bigcirc$) and 0.88 ($\square$).  Note the marked amplitudes of the noncrystallographic $\ell =5$ and $\ell=7$ invariants. 
}
\end{figure}

Figure \ref{figurezero} shows the dependence of $\langle Q^{2}_{\ell} \rangle$ on $\ell$ at temperatures 0.56 and 0.88.  The largest $Q_{\ell}$ are $Q_{6}$, $Q_{7}$, $Q_{12}$, and $Q_{5}$, in that order, followed by $Q_{11}$ and $Q_{10}$.  The significance of $Q_{7}$ does not appear to be anticipated in the published literature, because no space-filling structure has 7-fold (septahedral) symmetry, and the non-space-filling quasicrystalline symmetry that is usually discussed is the icosahedral symmetry, which emphasizes $\ell$ of 6, 10, and 12.  When the system is cooled from 0.88 to 0.55, $Q_{6}$, and to a lesser extent $Q_{7}$ and $Q_{12}$, increase.  All other $Q_{\ell}$ decrease with increasing temperature, the decreases in $Q_{4}$, $Q_{5}$, $Q_{8}$, and $Q_{9}$ being the most pronounced.  

\begin{figure}[bht]

\includegraphics{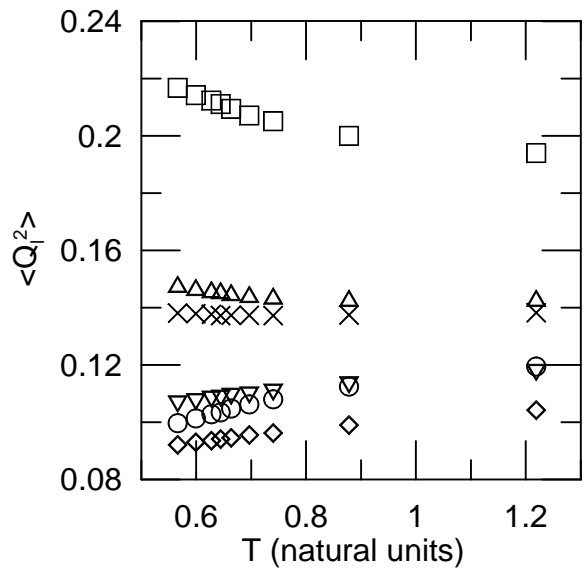}

\caption{\label{figure1}  Mean-square spherical invariants $\langle Q_{\ell}^{2} \rangle$ as a function of temperature for $\ell$ of 5($\bigcirc$), 6($\square$), 7($\vartriangle$), 10($\lozenge$), 11($\triangledown$), and 12 ($\times$). 
}
\end{figure}

Figure \ref{figure1} shows the temperature dependence of the six largest $\langle Q^{2}_{\ell} \rangle$.  The $\langle Q^{2}_{\ell} \rangle$ fall into three classes differing in their magnitudes and their temperature dependences.  $\langle Q^{2}_{6} \rangle$ is in the range 0.2-0.22; as the system is cooled from 0.88 to 0.56 it increases by almost 10\%. $\langle Q^{2}_{7} \rangle$ and $\langle Q^{2}_{12} \rangle$ are distinctly smaller than $\langle Q^{2}_{6} \rangle$, namely they are $\approx 0.14$ and increase modestly with decreasing temperature. $\langle Q^{2}_{5} \rangle$, $\langle Q^{2}_{10} \rangle$ and $\langle Q^{2}_{11} \rangle$ are $\approx 0.1-0.12$ at high temperature, and decrease by almost 10\% as T is reduced from 1.2 to 0.56.  $\langle Q^{2}_{10} \rangle$ is sometimes\cite{steinhardt1983a} associated with icosahedral ordering; we see it becomes smaller as temperature is reduced.  

We also determined the probability distributions, the likelihoods of determining particular values, for the $\langle Q^{2}_{\ell} \rangle$.  For each $\ell$, there are actually nine such functions, because one can measure a $\langle Q^{2}_{\ell} \rangle$ averaged over all central particles, or only with an A or a B particle at the origin. Furthermore, in evaluating $Q^{2}_{\ell}$ around a given center, one could consider all particles in a given shell, or consider only the A or the B particles in that shell. Limiting ourselves to an evaluation of $Q^{2}_{\ell}$ that includes all atoms in a given shell regardless of species, we found almost without exception that the probability distributions of the $Q^{2}_{\ell}$ are featureless, symmetric peaks whose widths at half height are typically half or 2/3 of the distribution's peak value.  The exception is $\langle Q^{2}_{9} \rangle$, as averaged over all central particles, which has a prominent side shoulder, because the $\langle Q^{2}_{9} \rangle$ for a central A and for a central B particle have very different average values.

To further explore how the $\langle Q^{2}_{\ell} \rangle$ are affected by the local composition we determined the conditional averages $\langle Q^{2}_{\ell} \rangle_{i,j,k}$, i.e., the value of $\langle Q^{2}_{\ell} \rangle$ for a given trio $i, j, k$.  Here $i$ is the identity of the central atom (A, B, All), while $j$ and $k$ are the number of A and B atoms, respectively, in the first shell; the calculation was made at $T = 0.56$.  The number of particles instantaneously found in the first coordination shell fluctuates a great deal, from as few as four to as many (with an A particle in the center) as fourteen.  When plotted as functions of $j$ and $k$, the $\langle Q^{2}_{\ell} \rangle_{i,j,k}$ fell into three families: (i) Functions such as $\langle Q^{2}_{5} \rangle_{A,j,k}$, $\langle Q^{2}_{6} \rangle_{B,j,k}$, and $\langle Q^{2}_{12} \rangle_{B,j,k}$ which decrease monotonically with increasing $j+k$, but are largely independent of $j-k$.  (ii) Functions such as $\langle Q^{2}_{5} \rangle_{B,j,k}$  that have a minimum at intermediate $j+k$ (for $\langle Q^{2}_{5} \rangle_{B,j,k} \rangle$, the minimum is at $j+k=8$), that have maxima at small and large $j+k$, and that are independent of $j-k$, and, finally, (iii) Functions such as $\langle Q^{2}_{6} \rangle_{A,j,k}$ that show a marked dependence on $j-k$, increasing at fixed $j+k$ as the number of B atoms in the shell is increased. 

A qualitative explanation for the trends in  $\langle Q^{2}_{\ell} \rangle_{i,j,k}$ with changing $j$ and $k$ may be suggested.  As $j + k$ is increased, the particles in the first coordination shell are obliged to pack more regularly rather than more randomly, at first making it more difficult for the instantaneous $Q^{2}_{\ell}$ to assume large values.  However, it is already quite difficult to pack 12 particles into the first coordination shell of a B particle, as witness that for a B center atom the first shell practically never includes more than twelve atoms, so in order to obtain $j+k$ = 12 the atoms in the first shell must be well-ordered, leading to a $\langle Q^{2}_{5} \rangle_{B,j,k}$ that increases at large $j+k$.  Finally, the A particles are larger than the B particles, making it easier to fill the first coordination shell as the number of B particles is increased, leading to a $\langle Q^{2}_{6} \rangle_{A,j,k}$ that increases with increasing $k-j$.

\begin{figure}
\includegraphics{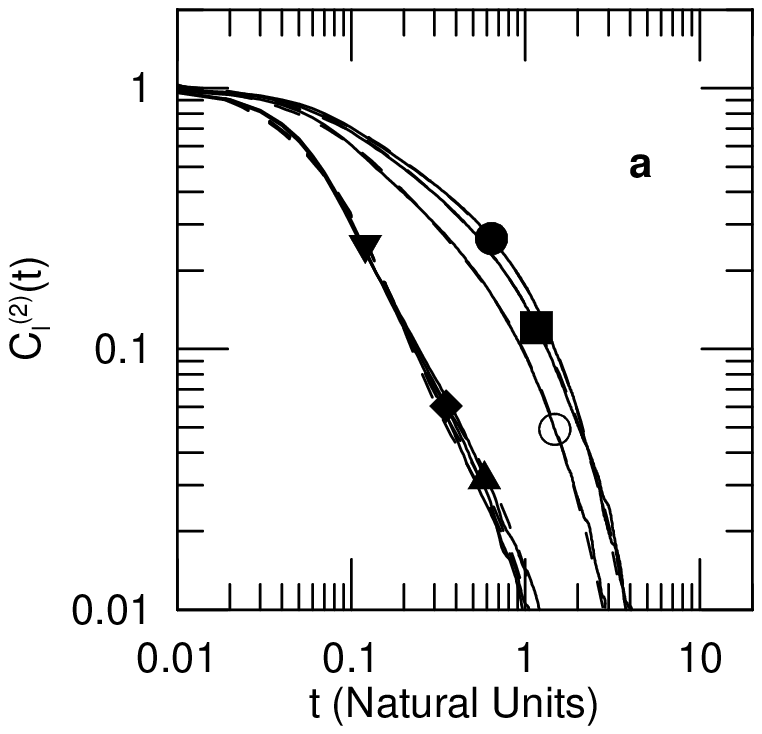}





\includegraphics{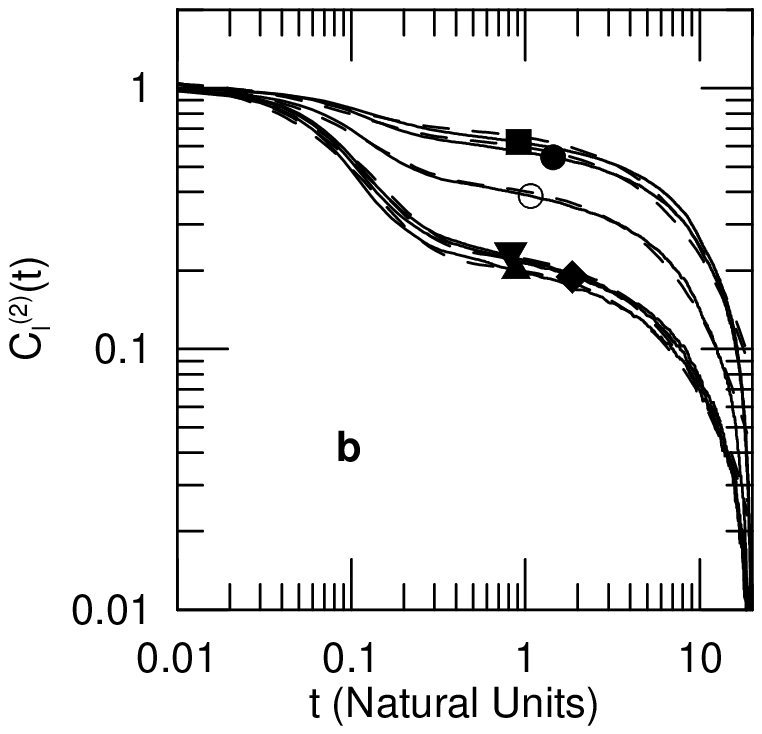}

\caption{\label{figure2}  Spherically invariant autocorrelation functions $C^{(2)}(t)$ formed from the $Y_{\rm LM}(t)$ for fixed $\ell$ of 5($\bigcirc$), 6($\bullet$), 7($\blacksquare$), 10($\blacktriangle$), 11($\blacklozenge$), and 12 ($\blacktriangledown$) at temperatures (a) 1.20 and (b) 0.56.
}

\end{figure}

In addition to examining the static correlations of the $Q_{\ell}$, we also examined their time autocorrelation functions. Preliminary studies found that all twelve $ C_{\ell}^{(2)}(t)$ relax on about the same time scale, so to reduce computational demands we focused on the $\ell$ having the largest magnitude, namely $\ell = 5, 6, 7, 10, 11,$ and $12$.  Measured correlation functions at the highest and lowest temperatures are found in Figure \ref{figure2}.  Figure \ref{figure2}a, at $T=1.20$, represents a liquid that is warmer than the measured\cite{whitford2005a} melting temperature $T_{m} \approx 1.1$.  At this temperature, the $\ell=10, 11, 12$ fluctuations are nearly indistinguishable, and decay the most rapidly, the $\ell =6$ and 7 correlations are the longest lived, and $C_{5}^{(2)}(t)$ has an intermediate life span.  Each curve was fit to the sum of a short- and a long-lived relaxation, as described below. The fits are represented by the dashed lines, which at higher temperatures are virtually indistinguishable from the data.   The slow relaxations extend the relaxation curves to longer times, but are not apparent in the Figure as long-time shoulders. 

In Figure \ref{figure2}b, ($T=0.56$, the lowest temperature studied), the slow-mode shoulders are clearly apparent, while the $C_{6}^{(2)}(t)$ and $C_{7}^{(2)}(t)$ functions are clearly distinct, $C_{6}^{(2)}(t)$ taking longer to decay. The amplitude of the slow mode of $C_{10}^{(2)}(t)$ is markedly less than the amplitudes of the slow modes of $C_{11}^{(2)}(t)$  and $C_{12}^{(2)}(t)$. The accuracy with which the $C_{\ell}^{(2)}(t)$ are described by the fitting process decreases with decreasing $T$, so that in Figure \ref{figure2}b the fitted curves (dashed lines) are no longer in near-perfect agreement with the measured data.

In order to obtain a quantitative description of the relaxations, each relaxation function was fit separately to a  form
\begin{equation}
    C_{\ell}^{(2)}(t) = A_{s} \exp(- \theta_{s} t^{\beta_{s}}) + A_{f} \exp(- \theta_{f} t^{\beta_{f}}).
    \label{eq:fitfunction}
\end{equation}
Here subscripts $s$ and $f$ denote the slow and fast modes, $A_{i}$ is an amplitude, $\theta_{i}$ is a decay pseudorate, and $\beta_{i}$ is a stretching exponent.  $A_{s}$ and $A_{f}$ were taken to be independent variables, rather than being constrained to sum to unity.  The $\beta_{i}$ were constrained to lie in the interval $[-1,1]$, but were free to vary within that interval (and in fact never entered the regime [-1, 0]), corresponding to processes that might be an exponential or a sum of exponential relaxations.  Removing the constraint on the $\beta_{i}$ sometimes gave very large values for $\beta_{i}$ (1.4 to 2), with a very small increase in fit accuracy.   When $\beta_{i} = 1$, the corresponding mode lifetime is $\theta_{i}^{-1}$.  In cases in which the relaxation was a stretched exponential in time, an average lifetime was assigned based on the exponential integral $\int_{0}^{\infty} dt \ \exp(-\theta t) = \theta^{-1}$, namely
\begin{equation}
      \langle \theta_{i}^{-1} \rangle \equiv \int_{0}^{\infty} dt \exp(-\theta_{i} t^{\beta_{i}}) = \theta_{i}^{-1/\beta_{i}} \Gamma(1+1/\beta_{i}), 
    \label{eq:averagelifetime}
\end{equation}
$\Gamma$ representing the Gamma function.

At low temperatures, best fits found pure exponential ($\beta =1$) relaxations for both modes.  At all temperatures, the shorter-lived mode was most often a pure exponential, though less so for $C_{6}^{(2)}$ and $C_{7}^{(2)}$ than for the other relaxations.  At higher temperatures, the slow mode is generally a stretched exponential, but as $T$ declines pure-exponential behavior is found.  The slow modes of $C_{10}^{(2)}$, $C_{11}^{(2)}$, and $C_{12}^{(2)}$ remain non-exponential to lower temperatures than their lower-$\ell$ counterparts.

Figures \ref{figure3}a and \ref{figure3}b show the lifetimes $\langle{\theta_{i}^{-1}}\rangle$ of the fast and slow modes for each $\ell$ at various temperatures.  For $\ell=10, 11, 12$, the fast mode lifetime increases gradually with decreasing temperature, but not by more than 30\%.  The slow mode lifetimes increase much more dramatically than the fast mode lifetimes, by factors of 12 to 20 over our temperature range.  Both fast and slow mode lifetimes tend to converge with decreasing $T$, so that at $T=1.2$ the lifetimes depend substantially on $\ell$, but at $T=0.56$ the lifetimes of each mode are nearly independent of $\ell$.  The fast-mode lifetimes of $C_{5}^{(2)}$, $C_{6}^{(2)}$, and $C_{7}^{(2)}$ show a novel temperature behavior, namely a non-monotonic temperature dependence, with $\langle \theta_{f}^{-1} \rangle$ climbing gradually to a peak at temperatures near 0.7, and then falling steeply as temperature is further reduced. 

Figures \ref{figure3}c and \ref{figure3}d give the amplitudes $A_{f}$ and $A_{s}$ for the fast and slow modes, respectively, against $T$.  The two amplitudes were fit independently, and consistently give $A_{f} + A_{s} \approx 1.1$.  The $\ell=10, 11,$ and $12$ relaxations are dominated by the fast mode, but for $\ell= 5, 6, 7$ the slow mode has, at the least, nearly the same amplitude to the fast mode.  The classes of temperature dependence seen above for the lifetimes are echoed by the amplitudes.  For $\ell = 10, 11, 12$, $A_{f}$ and $A_{s}$ are very nearly independent of $\ell$ and depend only weakly on $T$. The $\ell= 6$ and $7$ modes show a non-monotonic temperature dependence, the reversal in slope being near $T=0.7$.  For $\ell=5$, the amplitudes are very nearly independent from temperature.

Figures \ref{figure3}e and \ref{figure3}f give the fast and slow mode lifetimes as a functions of $\ell$ for various temperatures.  Figure \ref{figure3}e reinforces the prior description of mode behavior falling into three groups, namely $\ell=5$ (with a modest dependence of fast mode lifetime on $T$), $\ell=6$ and 7, with a wide range of $\langle \theta_{f}^{-1} \rangle$ with changing $T$, and $\ell=10, 11 ,12$, whose behaviors are very similar to each other.  Figure \ref{figure3}f presents the slow mode lifetime as a function of $\ell$.  While at each $\ell$ there is a strong dependence of $\langle \theta^{-1} \rangle$ on $T$, at fixed temperature the dependence of the slow mode lifetime on $\ell$ is much more modest.

\begin{figure}

\includegraphics{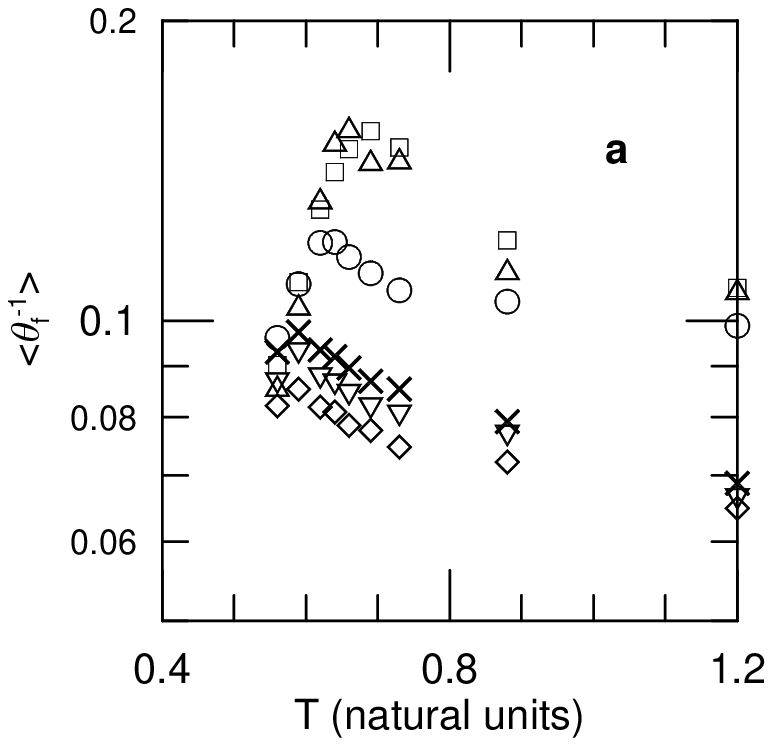}

\includegraphics{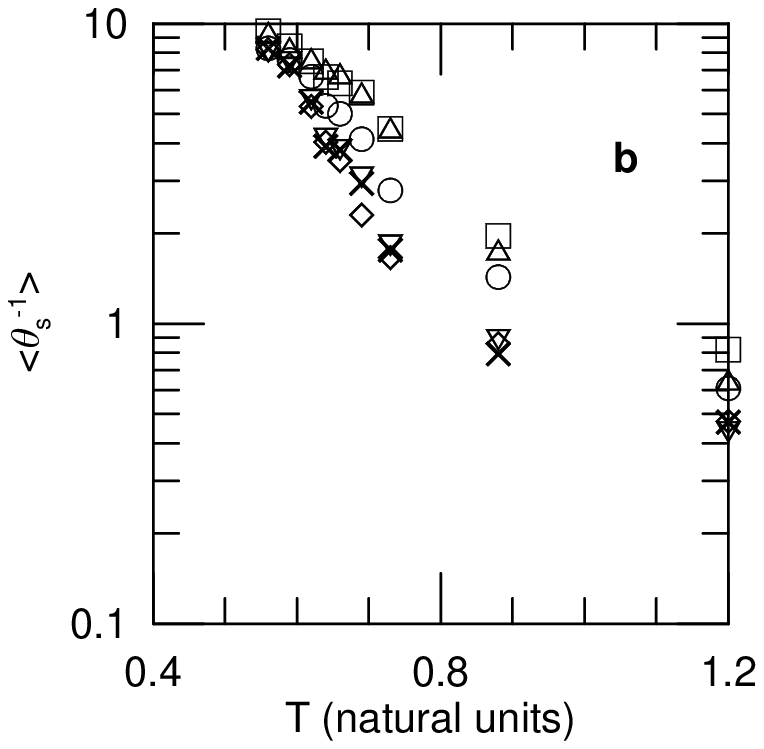}
\end{figure}

\begin{figure}

\includegraphics{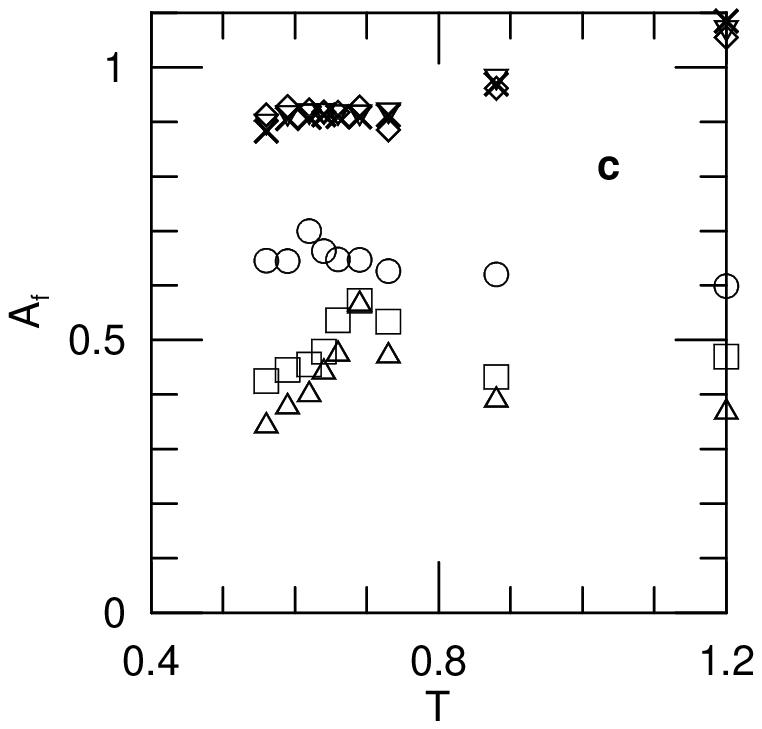}

\includegraphics{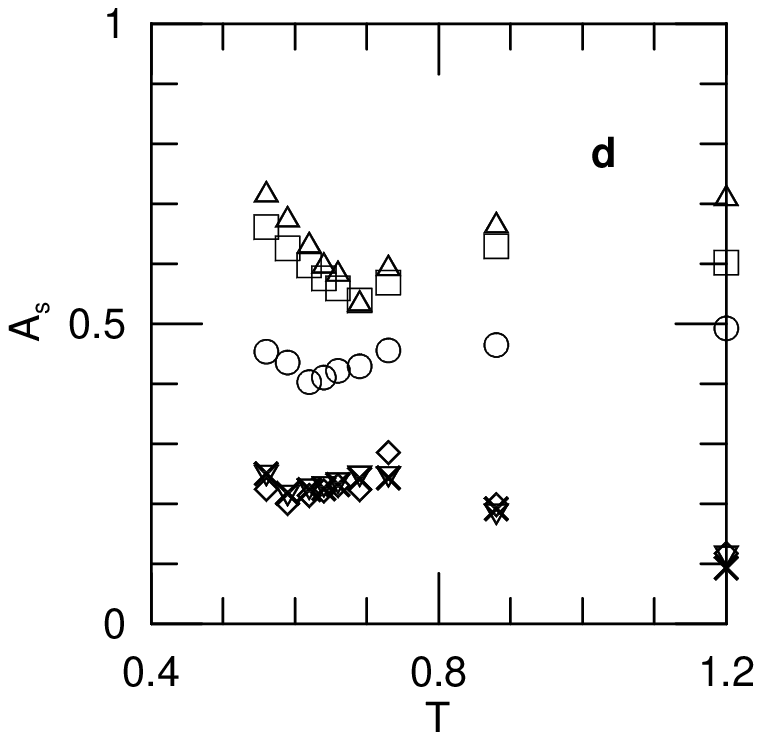}
\end{figure}

\begin{figure}

\includegraphics{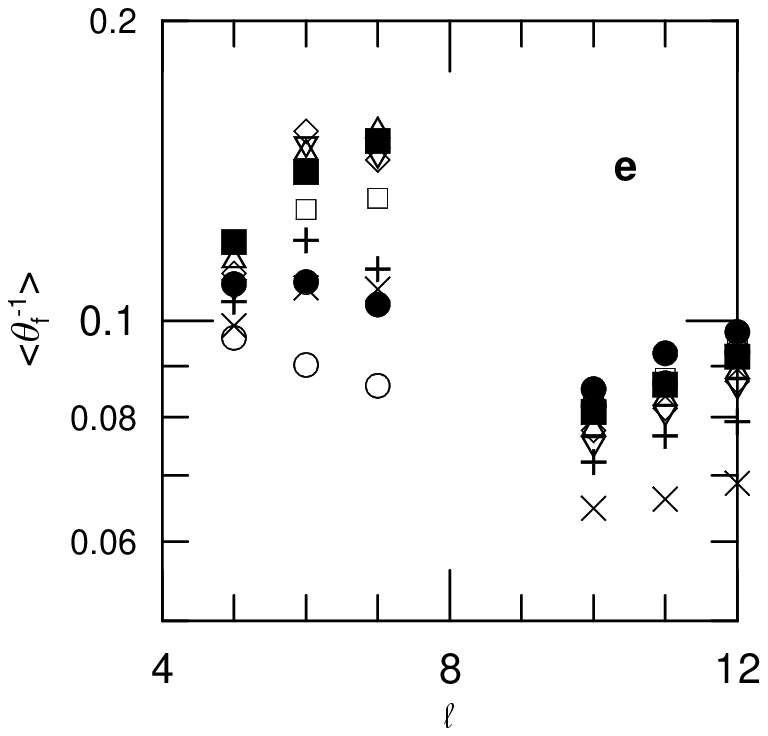}

\includegraphics{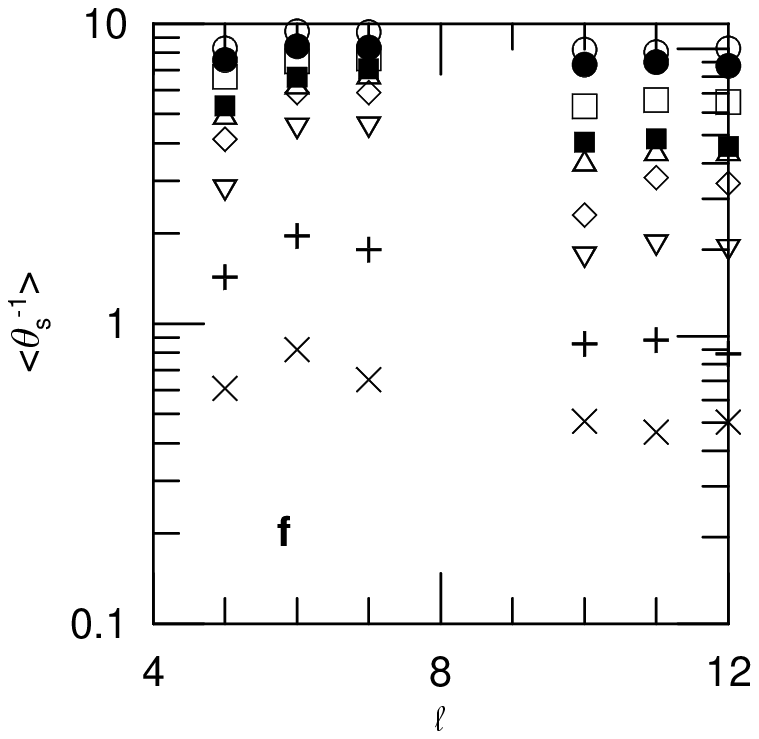}

\caption{\label{figure3}  Temperature dependence  of the (a) fast and (b) slow mode lifetimes $\langle \theta_{i}^{-1}\rangle$ and the (c) fast and (d) slow mode amplitudes $A$ for  fixed $\ell$ of 5($\bigcirc$), 6($\square$), 7($\vartriangle$), 10($\lozenge$), 11($\triangledown$), and 12 ($\times$). $L$-dependence of the (e) fast and (f) slow mode lifetimes at temperatures 0.56($\bigcirc$), 0.59($\bullet$), 0.62($\square$), 0.64($\blacksquare$), 0.66($\vartriangle$), 0.69($\lozenge $), 0.73($\triangledown$), 0.88($+$), and 1.20 ($\times$).
}

\end{figure}

We now consider the wavelet decompositions of the spherical harmonic densities.  We examined closely six values for $\ell$, three levels of decimation, and four filters, namely the mean-square outputs of the $CCC$, $DCC$, $DDC$, and $DDD$ filters, all at nine temperatures.  Because the system is rotationally invariant, the $DCC$, $CDC$, and $CCD$ filters are equivalent; we report the mean-square average of their outputs as $DCC$.  Figures \ref{figure7}a and \ref{figure7}b show the spherical harmonic densities as functions of temperature at decimation levels of 1 and 3, i. e., the densities averaged over an $S\times S \times S$ cube for $S$ of 2 and 8.  From Figure \ref{figure7}a, decimation level 1, as $T$ falls from 1.2 to 0.56 the smoothed densities for $Q_{5}$ and $Q_{10}$ decrease by nearly 30\%, the smoothed densities for $Q_{11}$ and $Q_{12}$ decrease slightly, and the smoothed densities for $Q_{6}$ and $Q_{7}$ increase, $Q_{7}$ by more than $Q_{6}$. On going to decimation level 2, which averages over 64 neighboring cubelets, the averaged $Q_{\ell}$ all decrease by nearly an order of magnitude, but the qualitative trends with temperature remain the same.  Between decimation level 2 and decimation level 3 (which averages over 512 neighboring basic cubelets), leading to Figure \ref{figure7}b, the smoothed $Q_{\ell}$ all fall  by approximately another factor of five, and the temperature dependences of $Q_{5}$ and $Q_{6}$  fade into the noise.  Figure \ref{figure7}c compares $Q_{5}$ and $Q_{7}$ at different decimations.   The temperature dependence of $Q_{5}$ decreases with increasing decimation, so that the first decimation of $Q_{5}$ decreases markedly with decreasing $T$, but the second and third decimations depend rather weakly on $T$.  In contrast, the temperature dependence of $Q_{7}$ increases with increasing order of decimation, so that the second decimation increases more swiftly with decreasing $T$ than does the first decimation, and the third decimation increases more swiftly with decreasing $T$ than does the second decimation.

\begin{figure}

\includegraphics{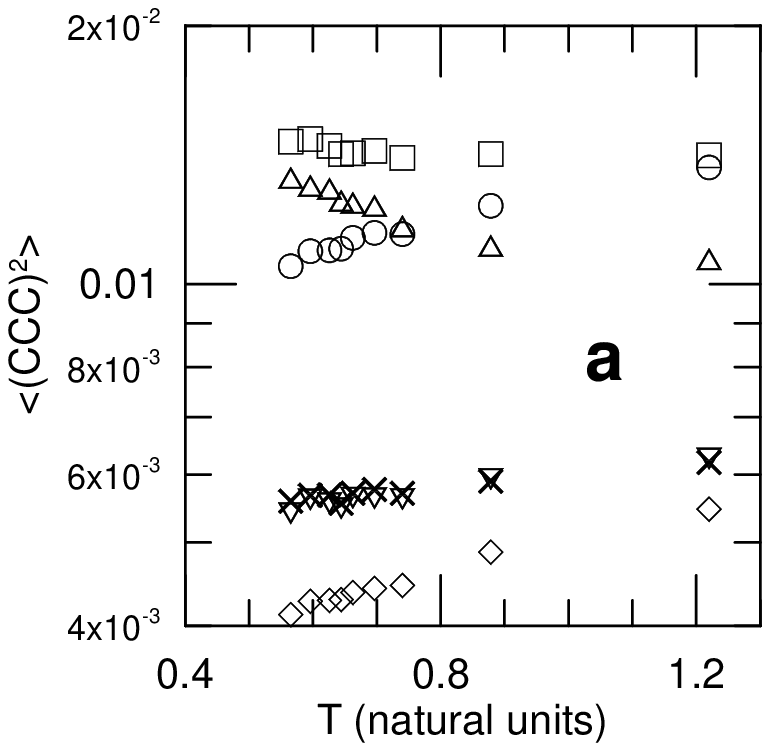}

\includegraphics{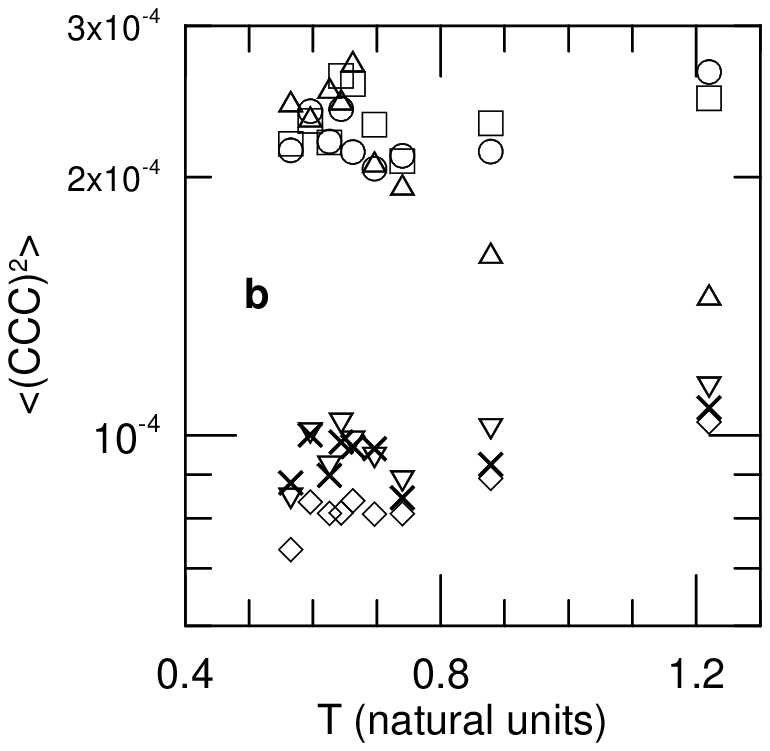}

\end{figure}

\begin{figure}

\includegraphics{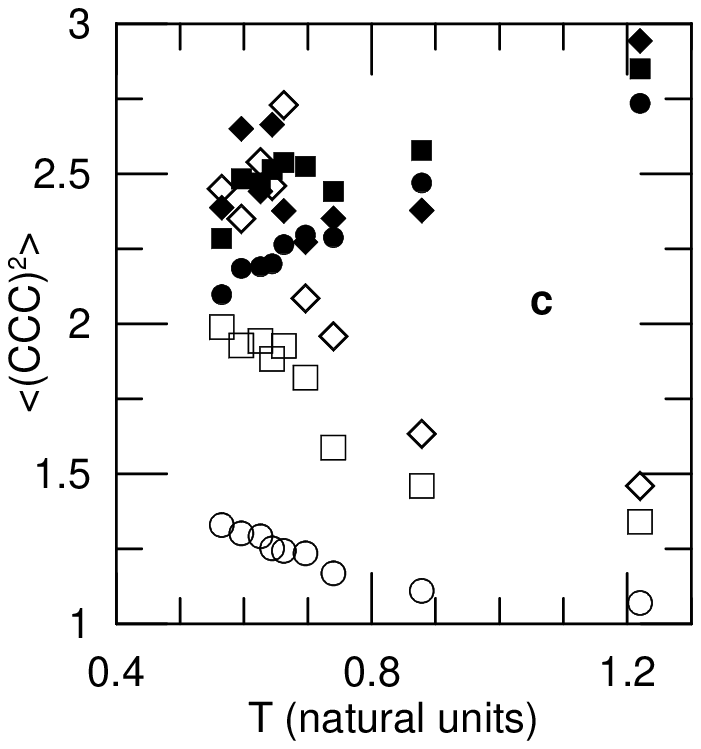}

\caption{\label{figure7}  Temperature dependence of the (a) first and (b) third Haar wavelet $\langle (CCC)^{2} \rangle$ decimations of the spherical harmonic component densities $Q_{\ell}$ for  $\ell$ using symbols of Figure \ref{figure3}, and (c) zeroth ($\bigcirc$), first ($\square$), and second ($\lozenge$) Haar wavelet decimations $\langle (CCC)^{2} \rangle$for $\ell = 5$ (filled points) and $\ell =7$ (open points) spherical harmonic densities.
}
\end{figure}

\begin{figure}

\includegraphics{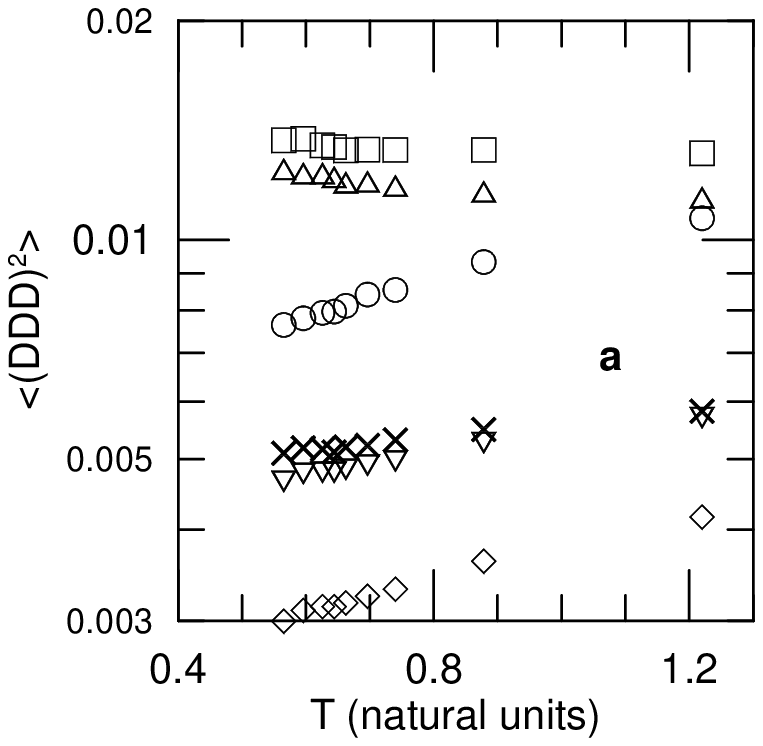}

\includegraphics{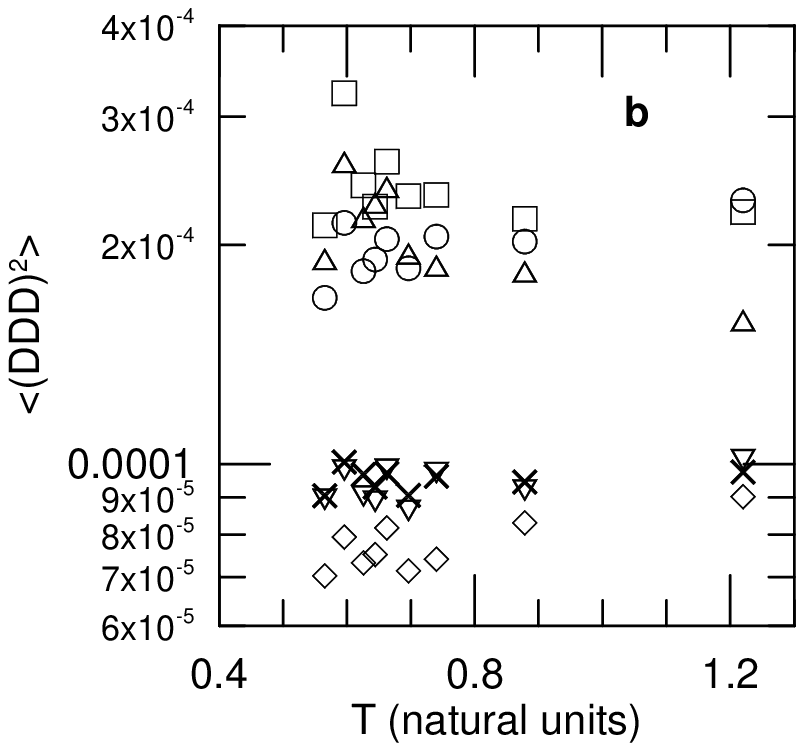}

\caption{\label{figure8}  Temperature dependence of the (a) first and (b) third Haar wavelet $\langle (DDD)^{2} \rangle$ decimations of the spherical harmonic component densities $Q_{\ell}$ for  $\ell$  using symbols of Figure \ref{figure3}.
}
\end{figure}

The numerical definitions of the $DCC$, $DDC$, and $DDD$ filters are quite different from each other, but at each decimation the plots of their respective outputs, as functions of $T$ for various $\ell$, are quite similar. The similarity arises because a change in temperature that increases the prominence of nominal vertices (as revealed by the $DDD$ filter) simultaneously increases the prominence of edges and faces on a volume (the $DDC$ and $DCC$ filters), all by commensurate amounts.  Figure \ref{figure8} shows the first and third $\langle (DDD)^{2}\rangle$ decimations as functions of $T$ for various $\ell$.  Just as the average $CCC$ components of $Q_{5}$ and $Q_{10}$ decrease as $T$ falls (cf. Fig \ref{figure7}), so also do the corresponding $\langle (DDD)^{2}\rangle$ components; similarly, just as the average $CCC$ component of $Q_{7}$ increases with falling $T$, so also do the corresponding $\langle (DDD)^{2}\rangle$ components.

How are the changes in local statics and dynamics correlated with other transport properties of the liquid?  One approach to determining the local resistance of the fluid to flow is to examine the time-dependent self-diffusion coefficient $D_{s}(t)$ of an atom of fluid, as determined from the mean-square displacement of the component atoms via
\begin{equation}
       \langle (\Delta {\bf r}(t) )^{2} \rangle = 6 t D_{s}(t).
       \label{eq:DsDrt}
\end{equation}
Here $\Delta {\bf r}(t)$ is the vector displacement of an atom during an interval $t$.  $D_{s}(t)$ may in turn be used to compute a formal time-dependent viscosity $\eta(t)$ as
\begin{equation}
    D_{s}(t) = \frac{ k_{B} T}{6 \pi \eta(t) R}.
    \label{eq:StokesEinstein}
\end{equation}
Here Boltzmann's constant $k_{B}$ is unity in natural units, $T$ is the absolute temperature in the same units, and $R$ is a particle radius.  If the mean-square displacement of a Lennard-Jones fluid atom is used to compute $D_{s}$ and thence $\eta$, the inferred viscosity is properly a microviscosity whose relationship to the orthodoxly-measured macroscopic viscosity is complex.

\begin{figure}

\includegraphics{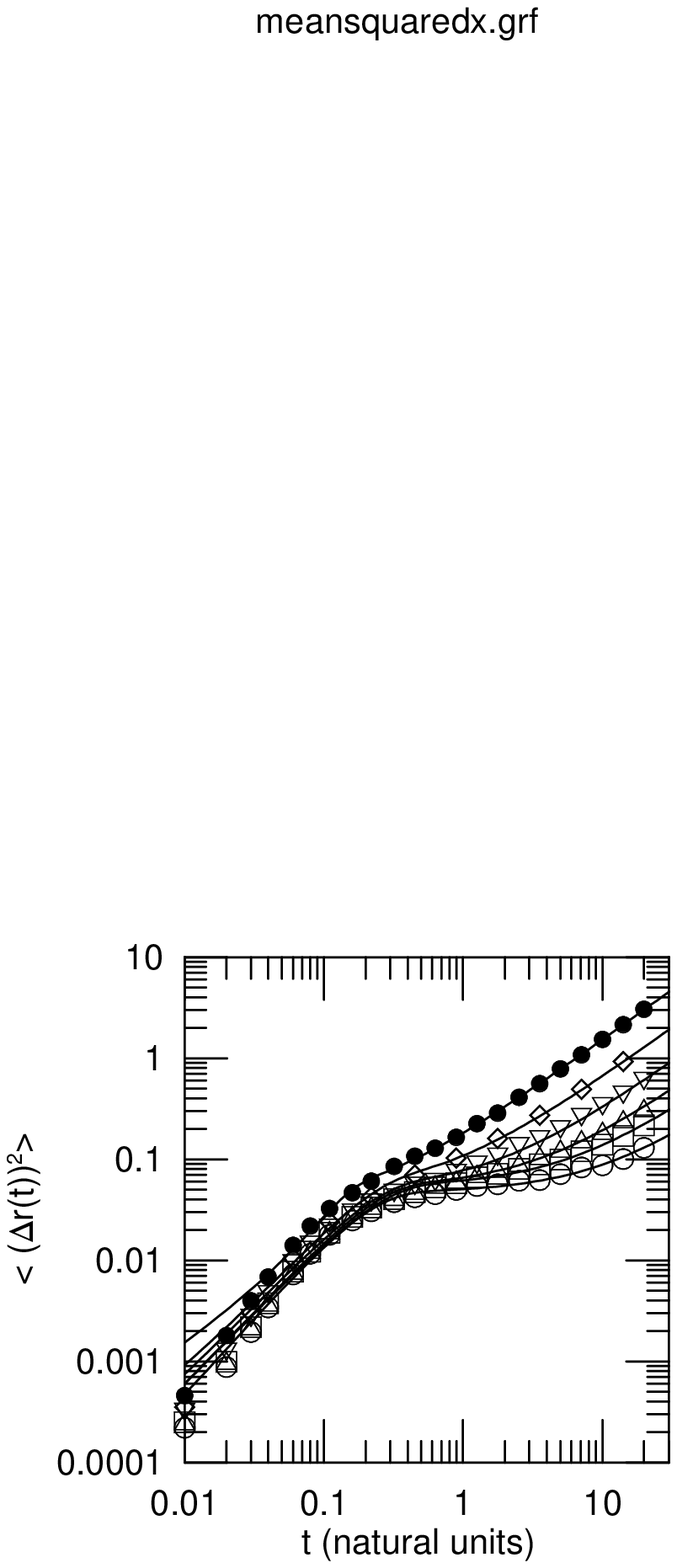}

\includegraphics{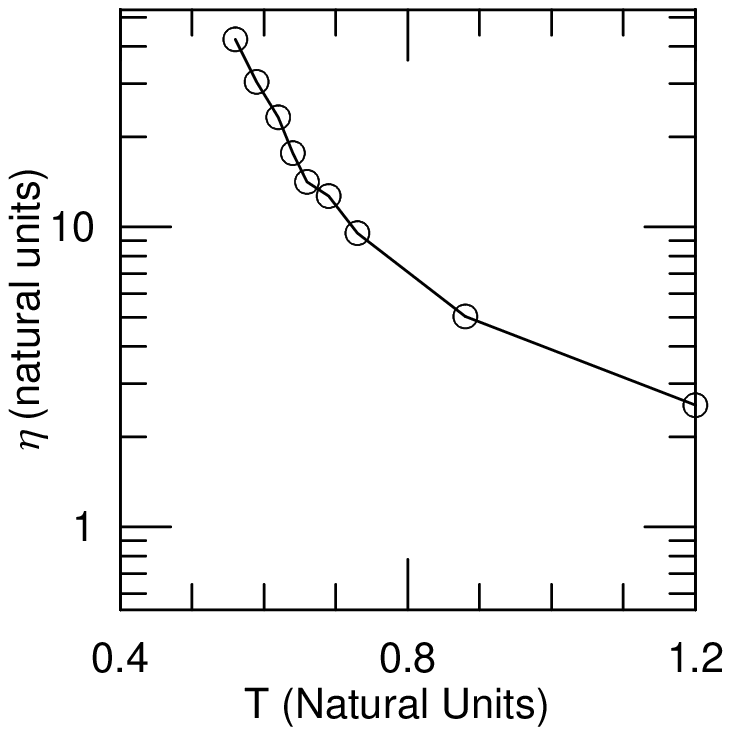}

\caption{\label{figure5}  (a) Mean-square displacement of a representative $A$ atom at temperatures (top to bottom) 1.20, 0.88, 0.73, 0.66, 0.62, and 0.56, and fits to eq \ref{eq:Drtfit}, and (b) long-time-limit solution microviscosity $\eta$ inferred from the inferred $D_{s}(\infty)$ with eqs \ref{eq:DsDrt}-\ref{eq:Drtfit} assuming $R\equiv \sigma_{AA} = 1$.
}

\end{figure}

We evaluated $\langle (\Delta {\bf r}(t) )^{2} \rangle$ for a randomly selected set of $A$ particles.  Representative determinations of $\langle (\Delta {\bf r}(t) )^{2} \rangle$ against $t$ appear in Figure \ref{figure5}a.  At short times, $\langle (\Delta {\bf r}(t) )^{2} \rangle$ has significant structure, but at long times the mean-square displacement tends to become linear in $D_{s}(\infty) t$.  The measured displacements were fit to
\begin{equation}
      \langle (\Delta {\bf r}(t) )^{2} \rangle = a \exp(-r t^{c}) + D_{s}(\infty)  t,
      \label{eq:Drtfit}
\end{equation}
with $a, b, c$, and $D_{s}(\infty)$ as fitting parameters.  Up to constants, the long-time microscopic shear viscosity is $\sim T/D_{s}^{-1}$, leading on identifying $R=\sigma_{AA}$ to the nominal microscopic viscosities seen in \ref{figure5}b.  Over the observed temperature interval, the microviscosity changes by a factor of 20, primarily at temperatures below 0.7.

\section{Discussion}

Here we have applied spherical harmonics and their invariants to study local orientational order in cold Lennard-Jones fluids.  Time correlation functions and wavelet decompositions were employed to examine the temporal and spatial persistence of the orientation order. We began with a $T=1.20$ fluid, a liquid slightly warmer than the apparent\cite{whitford2005a} crystalline melting point $T_{m} \cong 1.1$, and finished with a deeply-cooled $T=0.56$ vitrifying fluid.  Our particularly noteworthy findings include the importance of septahedral ($\ell = 7$) order, the appearance of slow modes in orientation relaxations $C^{(2)}_{\ell}(t)$, and the relationship between the orientation fluctuation amplitudes, orientation fluctuations lifetimes, and the microviscosity. 

As seen in Figure \ref{figure1}, the largest $\langle Q_{\ell}^{2} \rangle$ are those for $\ell$ of 6, 7, and 12.  
$\langle Q_{5}^{2} \rangle$ is next in importance, while $\langle Q_{11}^{2} \rangle$ is appreciably larger than $\langle Q_{10}^{2} \rangle$.  Earlier literature on spherical harmonic components of the density largely restricted itself to even $\ell$, and thus did not identify the importance of $\langle Q_{7}^{2} \rangle$ or $\langle Q_{11}^{2} \rangle$.  

The importance of $\langle Q_{6}^{2} \rangle$ is readily understood; it reflects a equatorial rings of six atoms surrounding the atom that is currently of interest.  The large size of $\langle Q_{7}^{2} \rangle$, which reflects the importance of seven-fold (septahedral) order, is unexpected and surprising.   In our other work\cite{whitford2005a}, the most stable crystal structure that we found for this system was body-centered-cubic, which leads to amplitudes for $\langle Q^{2}_{\ell} \rangle$ with $\ell$ even.  No crystal structure has fundamental seven-fold orientation order. Indeed, seven-fold symmetry of any sort is uncommon in nature.  A few complex organic compounds have distorted seven-fold symmetry\cite{crisma2001a,saviano2001a}, as by forming heptamers\cite{achsel2001a}.  Computer simulations of heptamer rings have been made\cite{wojciechowski2003a}, but no experimentally-observed quasicrystal has been reported to have seven-fold symmetry\cite{king2004a}.  The identifiable importance of unidecahedral ($\ell=11$) ordering, as observed here, is also unexpected.

The substantial sizes of the $\ell=10$ and $\ell=5$ components of $\langle Q_{\ell}^{2} \rangle$ are less surprising.
Proposals for the importance of quasicrystalline ordering in supercooled fluids \cite{kivelson1994a} have focused on icosahedral structures, which are not space-filling, but which are believed to be relatively stable for particles having Lennard-Jones interactions.  A local fluctuation in the form of a body-centered icosahedron contributes to $\langle Q_{10}^{2} \rangle$.  A pentagonal septode (a bifurcated half-icosahedron) and fused pairs of icosahedra also contribute to $\langle Q_{5}^{2} \rangle$.  

The temperature dependence of the $\langle Q_{\ell}^{2} \rangle$ speaks directly to the potential importance of different aspects of orientational order in the transition toward the glass.  As seen in Figure \ref{figure2}, as $T$ falls, so do $\langle Q_{5}^{2} \rangle$, $\langle Q_{10}^{2} \rangle$, and $\langle Q_{11}^{2} \rangle$; in contrast, as $T$ falls it is $\langle Q_{6}^{2} \rangle$, $\langle Q_{7}^{2} \rangle$, and $\langle Q_{12}^{2} \rangle$ that increase.  Thus, to the extent that orientational ordering with $\ell=5$ or $\ell=10$ reflects the formation of icosahedra and pentagonal septodes, the presence of those structures decreases as the system becomes more viscous.  Such a decrease in orientational ordering with decreasing $T$ is contrary to any hypothesis that vitrification is associated with icosahedral ordering that is enhanced by reducing the temperature.  The amplitudes that do increase are associated with $\ell$ of 6 and 12, which arise with conventional close packing of spheres, and $\ell=7$, which describes an elsewise-unspecified but certainly not space-filling septahedral ordering. 

Instead of proposing that $\langle Q_{10}^{2} \rangle$ or $\langle Q_{12}^{2} \rangle$ is uniquely characteristic of icosahedral order, one might instead note that the $\ell=11$ component have properties very similar to that of the $\ell=10$ and 12 components: their slow and fast mode amplitudes are very nearly equal at all $T$, their fast mode amplitudes are approximately equal, and their $\langle (CCC)^{2}\rangle$ and  $\langle (DDD)^{2}\rangle$ wavelet components have about the same size and temperature dependences. It might then alternatively be said, based on our sample, that all large-$\ell$ components have similar static and dynamic behaviors, and therefore the observed behavior of  $\langle Q_{10}^{2} \rangle$ and $\langle Q_{12}^{2} \rangle$ does not appear to arise from tendencies for icosahedral clustering.

Furthermore, there are considerable similarities between the behaviors of the $\ell = 5, 6, 7$ modes, similarities that make these modes systematically unlike the $\ell = 10, 11, 12$ modes in their behavior, in particular: The slow and fast modes are similar in amplitude, the fast mode lifetimes depend non-monotonically on $T$,  and for $\ell$ of 6 and 7 the mode amplitudes also depend non-monotonically on $T$, the slope reversals in all cases occurring near a $T$ of 0.7.  While the $\ell = 6, 10, 12$ amplitudes are associated with each other for icosahedral ordering, for the ordering actually found here it is the $\ell = 7, 6, 5$ amplitudes and separately the $\ell = 10, 11, 12$ modes that appear to be associated with ordering in the fluid.  Further study of higher-order spherical invariants might serve to clarify the nature of these orderings.

Wavelet decompositions of the local spherical harmonic densities also speak to the relative importance of the various $Q_{\ell}$.  Figure \ref{figure7}c shows the first three decimations of the low-frequency $\langle (CCC)^{2}\rangle$ parts of the $\ell=5$ and $\ell=7$ densities at various temperatures.  The amplitudes have been multiplied by constant factors, the same for all points of a given $\ell$ and decimation, so that fractional changes in the $\langle (CCC)^{2}\rangle$ as $T$ is changed are more readily compared.  It is important to emphasize that each decimation substantially reduces the $\langle (CCC)^{2}\rangle$, the reduction arising because with increasing decimation the  $\langle (CCC)^{2}\rangle$ represents an average of a fluctuation over a larger number of cubelets.  In the Figure, this reduction has been masked by the multiplicative displacement.  

For $\ell=5$, the decline in the $\langle (CCC)^{2}\rangle$ with decreasing $T$ is about the same for all three decimations, indicating that each decimation has about the same effect on $\langle (CCC)^{2}\rangle$ at all temperatures.  On distance scales $2^{m} \sigma_{AA}$, $m \in (1,3)$ probed by the decimations, the level of spatial coherence of $\langle Q_{5}^{2} \rangle$ as sampled by Haar low-pass wavelets is therefore independent of temperature.  In contrast, for $\ell=7$, the increase in $\langle (CCC)^{2}\rangle$ with decreasing $T$ becomes more dramatic as the decimation level is increased.  For the $m=1$ decimation, $\langle (CCC)^{2}\rangle$ changes by 20\% over the observed temperatures; for the $m=3$ decimation, the change in $\langle (CCC)^{2}\rangle$ is by nearly 2/3.
Recalling that each decimation greatly reduces $\langle (CCC)^{2}\rangle$,  the actual observed effect is that, with decreasing $T$ below approximately $T=0.7$, decimations become decreasingly effective in reducing  $\langle (CCC)^{2}\rangle$, the mean-square fluctuation in $Q_{7}$ as averaged over the region of support of a Haar smoothing wavelet.  Conversely, below $T=0.7$, the range over which fluctuations in septahedral ordering are correlated begins to increase, out to at least the distances $2\sigma_{AA}$ to $8\sigma_{AA}$ over which Haar smoothing wavelets have support. To the extent that orientational ordering is significant for glass formation, it is only the spherical harmonic components that increase with decreasing temperature that are plausibly contributing to vitrification, and these are the $\ell=6$ and $\ell=7$ components but not the $\ell=10$ or 12 components.

The $\langle (DDD)^{2}\rangle$ decompositions seen in Figure \ref{figure8} further emphasize the possible importance of $\langle Q_{6}^{2} \rangle$ and $\langle Q_{7}^{2} \rangle$ in vitrification.  For each decimation, $\langle (DDD)^{2}\rangle$ only increases with decreasing temperature for the $\ell =6$ and $\ell=7$ densities. For $\ell = 5, 10, 11$, and  $12$, at each decimation the high-frequency $\langle (DDD)^{2}\rangle$ components decrease or remain the same as $T$ is reduced.  

The temporal persistence of fluctuations in the $Q_{\rm LM}(t)$ is revealed by their time correlation functions, studied here from their spherical invariants $C^{(2)}_{\ell}(t)$.  The $C^{(2)}_{\ell}(t)$ all have a bimodal structure, with a rapidly-decaying mode whose lifetime is nearly independent of $T$, and a slowly-decaying mode whose lifetime increases dramatically with decreasing $T$.  The slow mode amplitude is quite weak at $T=1.2$, is readily apparent at $T=0.88$, and is much more prominent at lower temperatures.  

The fast mode has a relaxation time near 0.1 in natural units, relaxation times extending over no more than a factor of 2, over the full range of $T$ and $\ell$ that we examined. The relaxation time of the fast modes is very close to the periods of onset--time from initial zero to first maximum-- of the correlation functions $\langle Y_{1} P(\tau) \rangle$ and  $\langle Y_{2} K_{z}(\tau) \rangle$.  It might therefore be proposed that the fast mode corresponds to an initial period during which particle motions are highly correlated with aspects of initial particle positions.

The slow mode lifetimes increase from $0.7 \pm 0.2$ at T=1.20 to $\approx 10$ at $T=0.56$.  At higher temperatures, between $\ell=6$ and $\ell=12$, $\langle \theta_{s}^{-1} \rangle$ varies by by more than a  factor of two, the larger-$\ell$ correlation functions having the shorter lifetimes.  As temperature is reduced, the  $\langle \theta_{s}^{-1} \rangle$ converge toward a common $\ell$-independent value. The temperature at convergence is certainly not greatly different from the Mode Coupling Theory critical temperature estimated for this system by Sastry, et al.\cite{sastry1999a}. We did not reach temperatures low enough to clarify if the relaxation times continue to keep a common value at temperatures colder than convergence, or if they again diverge from each other as $T$ is further reduced.  For $\ell = 10, 11, 12$, the increase in $\log(\langle \theta_{s}^{-1} \rangle)$ with falling $T$ is more rapid than linear in $T$, so that $\langle \theta_{s}^{-1} \rangle$ for these modes might be diverging at some positive non-zero temperature.  In the direction of higher temperatures, extrapolation of the slow and fast mode lifetimes predicts that the lifetimes become equal near $T \approx 2$.  This temperature has had significance previously: In our previously-published\cite{whitford2005a} study of this system, a lower-temperature local structure formation was identified from the radial distribution function as appearing near $T=2$ and becoming more prominent as $T$ is reduced.

The $\ell$-dependence of $\langle \theta_{s}^{-1} \rangle$ suffices to prove that the relaxation of the $C_{\ell}^{(2)}(t)$ cannot arise from the rotational diffusion of quasi-rigid evanescent structures, namely:  If a rigid structure is performing rotational diffusion, the relaxation of its $\ell^{\rm th}$ spherical harmonic component will follow $\exp(- \ell (\ell+1) D_{\theta} t)$, where $D_{\theta}$ is the rotational diffusion coefficient\cite{berne}. The relaxation time is inversely proportional to $\ell(\ell+1)$, so for $\ell$ between 5 and 12 the time would change more than fivefold. The fast mode properties are inconsistent with rotational diffusion: $\langle \theta_{f}^{-1} \rangle$ varies with $\ell$ by less than a factor of two; furthermore, the fast mode of $C_{12}^{(2)}(t)$ is longer-lived than the same mode of $C_{11}^{(2)}(t)$, which is in turn longer-lived than the fast mode of  $C_{10}^{(2)}(t)$, an order inverse to that expected from rotational diffusion.  The slow mode properties are equally inconsistent with rigid-body rotational diffusion.  Even at $T=1.2$ the change in $\langle \theta_{s}^{-1} \rangle$ with $\ell$ is by less than a factor of three, is nearly independent of $\ell$ for $\ell$ in the range 10-12, and at low temperatures is very nearly independent of $\ell$ for all $\ell$ studied, entirely inconsistent with the rotational diffusion expectation that the rotational diffusion time should scale as $\ell (\ell+1)$. 

\begin{figure}

\includegraphics{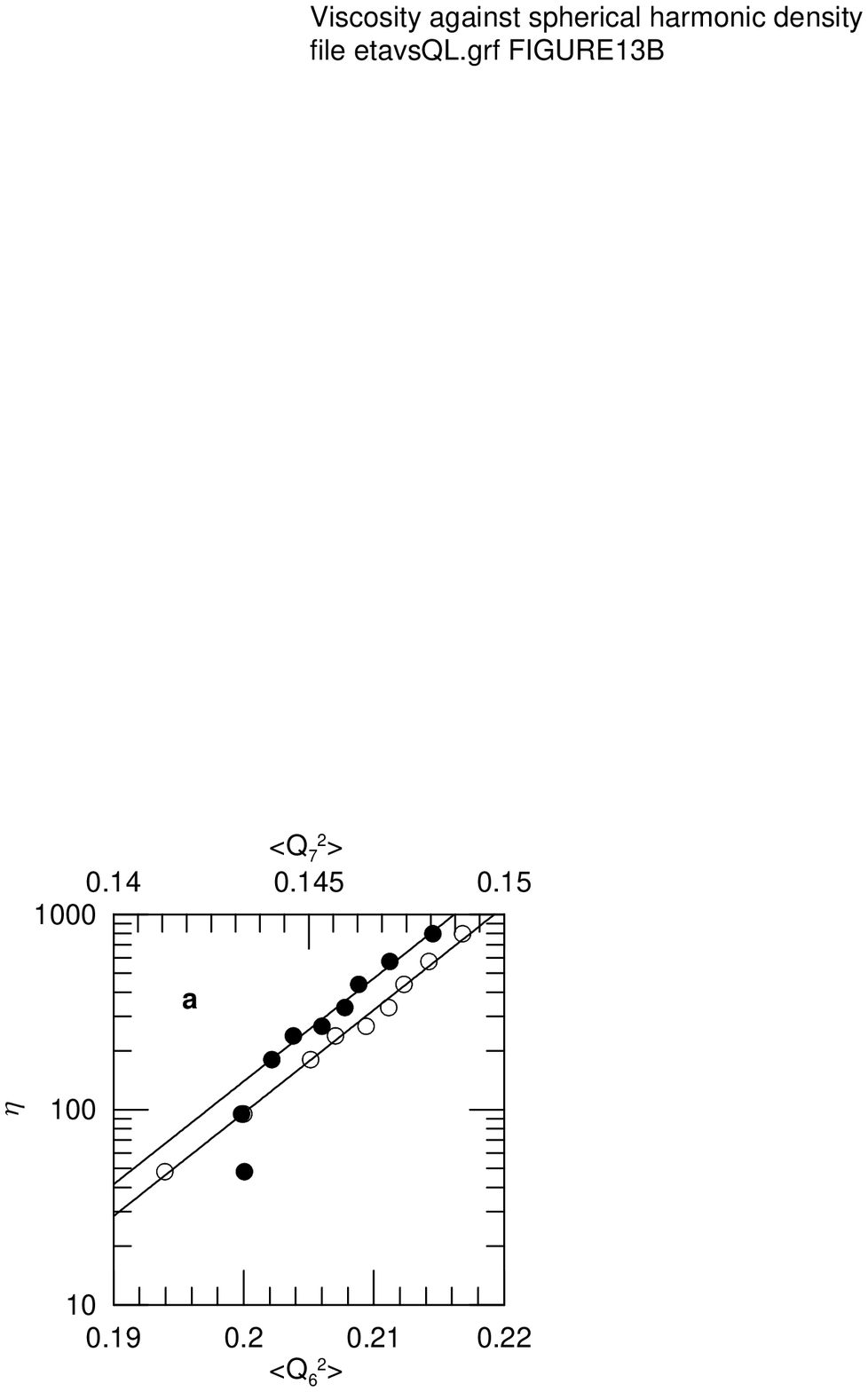}

\includegraphics{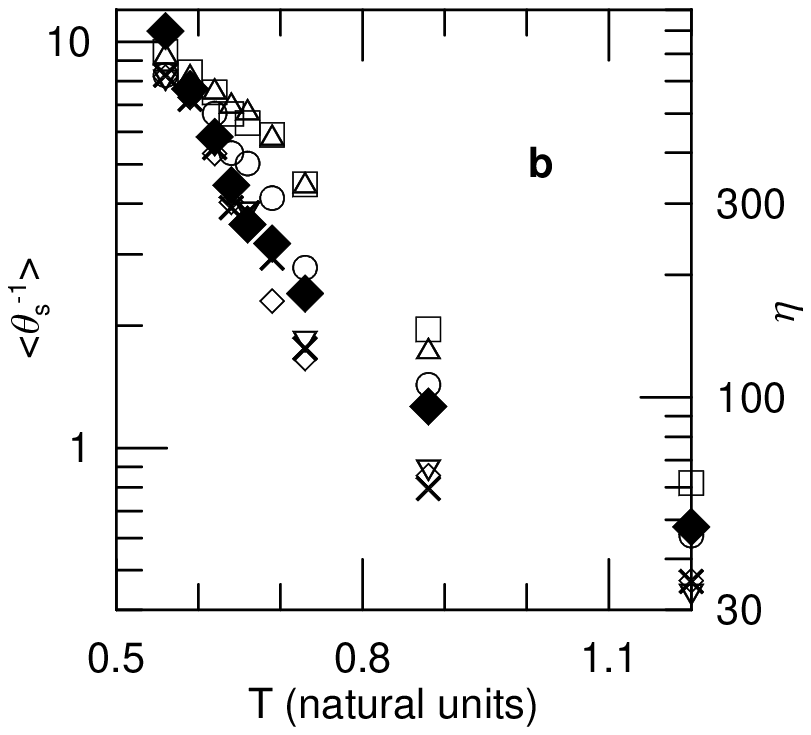}

\caption{\label{figure13}  (a) Viscosity $\eta$ from long-time self-diffusion coefficient $D_{s}(\infty)$ plotted against $\langle Q_{6}^{2} \rangle$ ($\bigcirc$) and $\langle Q_{7}^{2} \rangle$ ($\bullet$) and (lines) fits to simple exponentials.  (b) Viscosity (right axis, $\blacklozenge$) and slow-mode relaxation times (left axis, points as per Figure \ref{figure3}) as functions of temperature.
}
\end{figure}

Figure \ref{figure13}a demonstrates that the apparent microviscosity of the liquid, as inferred from the particle self-diffusion coefficient, is correlated with the orientation amplitudes $\langle Q_{6}^{2} \rangle$ and $\langle Q_{7}^{2} \rangle$.  As indicated by the straight lines, for both amplitudes one sees
\begin{equation}
      \eta =  \eta_{o} \exp(a \langle Q_{\ell}^{2} \rangle).
      \label{eq:etaQLreln}
\end{equation}
Here $\eta_{o}$ is $\eta$ at the perhaps-inaccessible $\langle Q_{\ell}^{2} \rangle=0$, and $a$ is the slope seen in Figure \ref{figure13}a.  Is this exponential correlation credible?  There are other systems in which the viscosity increases more or less exponentially with the concentration of the component responsible for the increase in viscosity. For example, in many polymer solutions over a wide polymer concentration range the viscosity of a polymer solution increases more or less exponentially in the polymer concentration\cite{polymereta}. The correlation is therefore not improbable on its face. From Figure \ref{figure13}b, the temperature dependence of the solution microviscosity is also not that different from the temperature dependences of the slow mode lifetimes. Over a shared range of decreasing temperatures, $\eta$ and $\langle \theta_{s}^{-1} \rangle$ each increase almost twenty-fold. 

There is a substantial literature on simulations of Lennard-Jones fluids, and a limited literature on spherical harmonic decompositions of particle densities.  We first note several simulations that have found phenomena seemingly related to those studied here.  We then turn to historical studies of spherical harmonic decompositions.

For a system with the composition and potential energy studied here, Sastry, et al.\cite{sastry1999a} found that the self intermediate scattering function $F_{sA}(q,t)$ of their $A$ particles gains a strong slow mode as $T$ is taken from 0.88 down to 0.73. The slow mode in $F_{sA}$ is a stretched exponential in time, whose stretching exponent $\beta$ is $\approx 0.9$ at temperatures 0.88 and above; at temperature 0.73 and below $\beta$ declines rapidly, reaching $\beta \approx 0.55$ by temperature 0.59. The mode coupling critical temperature for the system was estimated to be $T_{c}=0.592$. Sastry, et al.\ also examined the distribution of particle displacements at fixed time and various temperatures, comparing distributions at the time at which the mean-square particle displacement was unity.  The distribution of displacements was nearly Gaussian above $T \approx 0.63$; at lower temperatures the distribution broadens and bifurcates.  Between temperatures 0.88 and 0.59, the lifetime of the slow mode in $F_{sA}(q,t)$ increases by two orders of magnitude, considerably more than the increase that we found in $\langle \theta_{s}^{-1} \rangle$ over the same range of temperatures.  Nonetheless, the slow modes we find in $C_{\ell}^{(2)}(t)$ are clearly accompanied by slow modes found in more familiar time correlation functions.

A series of studies on Lennard-Jones fluids, e.g., by Donati, et al.\cite{donati1999a}, have found that relatively mobile and relatively immobile particles separately tend to be found in clusters.  The clusters of mobile particles are long and ribbonlike; clusters of immobile particles are more compact. Donati, et al.\ present evidence that the liquid structure is much more ordered near an immobile particle than it is near a mobile particle, the additional ordering persisting out to a radius $\approx 4 \sigma_{AA}$.  The distance over which immobile particles are correlated does not appear to increase much as temperature is reduced.  On the other hand, the relative positions of highly mobile and highly immobile particles tend to be anticorrelated, the anticorrelation having a length scale that increases with decreasing $T$.  From our wavelet decompositions, we find that the implicit range of some correlation functions, e.g., $\langle Q_{5}^{2} \rangle$, does not appear to grow with falling temperature, just as Donati, et al., found that clusters of immobile particles do not appear to grow at $T$ is reduced.  On the other hand, we found from wavelet decompositions evidence that the range over which fluctuations in $\langle Q_{7}^{2} \rangle$ are correlated appears to increase as temperature is reduced, just as Donati, et al. found that range of anticorrelations between mobile and immobile particles grew as $T$ was reduced.  It will be interesting in future work to examine correlations between the details of local orientational order and the local mobility of particles of interest.

Berthier\cite{berthier2005a} determined a temperature-dependent range $\xi$ for spatial correlations  of particle mobilities, showing that $\xi$ determines the distance scale at which a relaxation time $\tau$ for particle motion decouples from the macroscopic viscosity $\eta$, so that at lower temperatures $\tau \eta$ becomes temperature-dependent.  Berthier proposes that his simulations show that the properties of supercooled liquids, and thus glass formation, are influenced by extensive spatial correlations, as seen previously in studies of critical phenomena.  This proposal is said by Berthier\cite{berthier2005a} to be contrary to some models for glass formation. Figure \ref{figure13} shows a phenomenon similar to that discussed by Berthier, namely that the microviscosity (i.e., the self-diffusion coefficient) tracks accurately the degree of static orientational ordering for hexahedral and septahedral ordering, and less precisely tracks the orientation fluctuation correlation times from $C_{\ell}^{(2)}(t)$.  Our results differ from Berthier's in that our results speak directly to localized ordering, involving the first coordination shell, rather than to the spatially extended ordering seen in his analysis.  However, our wavelet decompositions indicate that with declining temperature the septahedral ordering, in addition to becoming more pronounced, is also increasing the range of its spatial correlations.  Our analysis does not show whether the relationship between the range increase we observe and the change in microscopic viscosity is causal or concomitant.

The notion that local ordering in fluids may be noncrystallographic was proposed by Frank\cite{frank1952a}, who noted that local icosahedral packing of Lennard-Jones particles was energetically preferred to hexagonal close-packed or face-centered cubic arrangements.  Steinhardt, et al.\cite{steinhardt1983a} proposed to characterize local order in a fluid by making spherical harmonic decompositions of the local density of "bond" (near-neighbor vector) directions.  They emphasized the importance of spehrical invariants such as $Q_{\ell}^{2}$. Icosahedral ordering contributes to invariants with $\ell$ of 6 and 10, while cubic ordering has little invariant with $\ell=10$ but substantial invariant components with $\ell$ of 4 and 8. Steinhardt, et al., report simulation results for even $\ell \in (2,10)$, as well as for more complicated spherical invariants, concluding that significant icosahedral and some cubic ordering is present.  Tomida and Egami\cite{tomida1995a} extended this approach by considering $Q_{\ell}^{2}$ as computed for bonds centered on atoms, rather than spatial regions.  Clusters centered on atoms provide a natural definition of local clustering. Their simulations used a potential suited for liquid iron, not the Lennard-Jones potential used here.   They also found the probability distribution for the $Q_{\ell}^{2}$.  Clusters for which $Q_{6}^{2}$ is large, as expected for icosahedra, were found by Tomida and Egami to have values for $\langle Q_{2}^{2} \rangle$, $\langle Q_{4}^{2} \rangle$, $\langle Q_{8}^{2} \rangle$, and $\langle Q_{10}^{2} \rangle$ expected for icosahedra.  However, the hypothesis of icosahedral ordering in fluids is not without controversy, as witness the careful analysis of Stillinger and LaViollete\cite{stillinger1986a,laviollete1990a}.

The results here do not entirely agree with the literature, in part because we asked a somewhat different set of questions.  In particular, we are not aware of prior measurements of $\langle Q_{7}^{2} \rangle$ in this or similar systems, so prior studies would not readily have noticed the possible importance of septahedral ordering.  Prior studies measured $\langle Q_{10}^{2} \rangle$, finding that it was non-zero.  Icosahedral, but not cubic, order leads to a large $\langle Q_{10}^{2} \rangle$, leading to the conclusion that icosahedral clustering is likely important in supercooled fluids.  We measured not only $\langle Q_{10}^{2} \rangle$ but also $\langle Q_{11}^{2} \rangle$ and $\langle Q_{12}^{2} \rangle$, finding that these three spherical harmonic components are very similar to each other in their magnitudes, relaxation times, and temperature dependences. $Q_{11}$ is not associated with icosahedral structures, so the observation that these three large-$\ell$ components have very similar behaviors suggests that the observed behavior of $Q_{10}$ and $Q_{12}$ is some general large-$\ell$ effect, and not some consequence of icosahedral structure formation that would be expected to contribute to $Q_{10}$ but not to $Q_{11}$.

Kivelson, et al.\cite{kivelson1994a} have given a detailed model for glass formation in liquids,  based on the formation of frustration-limited icosahedral clusters, starting at a cluster melting temperature significantly warmer than the melting temperature of the (unfrustrated) crystalline phase.  In our previous paper, we proposed that we saw evidence for such a model, in the form of local, growth-limited clusters that appeared below a temperature $T \approx 2$ that is well above the crystalline melting temperature $T_{\rm cm} \approx 1.1$ that we determined at our density. One of us\cite{phillies2004a} has assembled evidence indicating that the so-called neutral polymer slow mode observed in light scattering spectra of some but not other nondilute solutions of neutral polymers provides experimental evidence for Kivelson's model. In comparing results here with the proposals of Kivelson, et al., it is important to note that nothing in the Kivelson model relies on the growth-frustrated clusters being icosahedral.  The non-space-filling septahedral ($Q_{7}^{2}$) ordering for which evidence is found here is entirely consistent with the Kivelson model's demands for a growth-limited spatial structure.

\end{document}